\documentclass[11pt,twoside]{article}
\usepackage{asp2010}
\aspcpryear{2010}
\aspvoltitle{UP: Have Observations Revealed a Variable Upper End of
  the Initial Mass Function?}

\resetcounters

\begin{document}

\title{The Role of the IGIMF in the chemical evolution of the solar neighbourhood}
\author{Francesco Calura$^1$, Simone Recchi$^2$, Francesca Matteucci$^3$, Pavel Kroupa$^4$
\affil{$^1$Jeremiah Horrocks Institute for Astrophysics and Supercomputing, University of Central Lancashire, Preston PR1 2HE, United Kingdom}
\affil{$^2$Institute of Astronomy, Vienna University,
                T\"urkenschanzstrasse 17, A-1180, Vienna, Austria}
\affil{$^3$Dipartimento di Fisica, Sezione di Astronomia, Universit\`a di Trieste, via G.B. Tiepolo 11, 34131, Trieste, Italy}
\affil{$^4$Argelander Institute for Astronomy, Bonn University,
            Auf dem H\"ugel 71, 53121 Bonn, Germany}}

\begin{abstract}
The integrated galactic initial mass function (IGIMF) is computed from the combination of the
stellar initial mass function (IMF) and the embedded cluster mass function, described by a power law with
index $\beta$. 
The result of the combination is a time-varying IMF which
depends on the star formation rate. 
We applied the IGIMF  formalism to a
chemical evolution model for the solar neighbourhood and compared the
results obtained by assuming three possible values for $\beta$ with the
ones obtained by means of a standard, well-tested, constant IMF. 
In
general, a lower absolute value of $\beta$ implies a flatter IGIMF, hence
a larger number of massive stars,  
higher Type Ia and II supernova rates, higher
mass ejection rates 
and higher  [$\alpha$/Fe] values at a given metallicity.  
Our suggested fiducial value for $\beta$ is 2, since with this value we can 
account for most of the local observables. 
We discuss our results in a broader perspective, with some implications regarding the possible universality of the IMF and the importance of the 
star formation threshold. 
\end{abstract}

\section{Introduction}
The initial stellar mass function (IMF) is of primary importance 
in galactic chemical evolution models. 
The IMF regulates the relative fractions of stars of different masses, hence their relative 
contribution to the chemical enrichment of the interstellar medium (ISM) is tightly related to this quantity. 
For this reason,  the analysis of abundance ratios in galaxies  
may allow one to put robust 
constraints on  both the normalization and the slope of the IMF (Recchi et al. 2009; Calura et al. 2010). \\
The Solar Neighbourhood (S. N. hereinafter) can be considered the most valuable environment to achieve constraints 
on the main parameters regulationg chemical evolution models, 
since it is definitely the best studied Galactic environment and 
many observational investigations devoted to its study 
provide us with a large set of observables against which models can be tested. 
These observables include diagrams of abundance 
ratios versus metallicity, which are particularly useful when they involve two elements synthesised 
by stars on different timescales. 
An example is the 
[$\alpha$/Fe] vs [Fe/H] diagram, since $\alpha$ elements are produced mostly by massive stars ($m>8 M_{\odot}$) on very 
short ($<0.03$ Gyr) timescales, 
while type Ia supernovae (SNe) produce most of the Fe on timescales ranging from 0.03 Gyr up to one Hubble time (Matteucci 2001). 
This diagnostic is a strong function of the IMF, but 
depends also on the assumed star formation (SF) history (Matteucci 2001; Calura et al. 2009). 
Another fundamental constraint is the stellar metallicity distribution (SMD), 
which depends mainly on the IMF and on the infall history (hence on the star formation history) 
of the studied system. 
Another example of a useful diagnostic test for the IMF 
is the present-day mass function, i.e. the mass function of living stars observed now in the Solar 
Vicinity (Elmegreen \& Scalo 2006). \\
The integrated galactic initial mass function (IGIMF) 
originates from the combination of the stellar IMF within each star cluster and 
of the embedded cluster mass function (CMF).
It relies on the observational evidence that 
small clusters are more numerous in galaxies and that 
the most massive stars tend to
form preferentially in massive clusters (Weidner \& Kroupa 2006). The IGIMF is star-formation dependent, 
hence it is time-dependent and its evolution with time is sensitive to the star formation history of the environment. \\
In this paper, we use all the local observables 
to study the IGIMF and its effects on the 
chemical evolution of the solar neighbourhood. 
The results obtained with the IGIMF are compared to those obtained with a non-star-formation dependent (hence constant in time), 
fiducial IMF. The aim is to derive some contraints on the main unknown parameter of the IGIMF, i.e. the 
index $\beta$ of the power law expressing the embedded CMF.\\
This paper is organized as follows. In Section 2 we present a description of the theoretical scenario behind the IGIMF. In Section 3 we describe 
the 
chemical evolution model for the Solar Neighbourhood. 
In Sect. 4 we present our results and finally in Sect. 5 some open problems regarding the IMF are discussed and some conclusions are drawn.

\begin{figure}
\plottwo{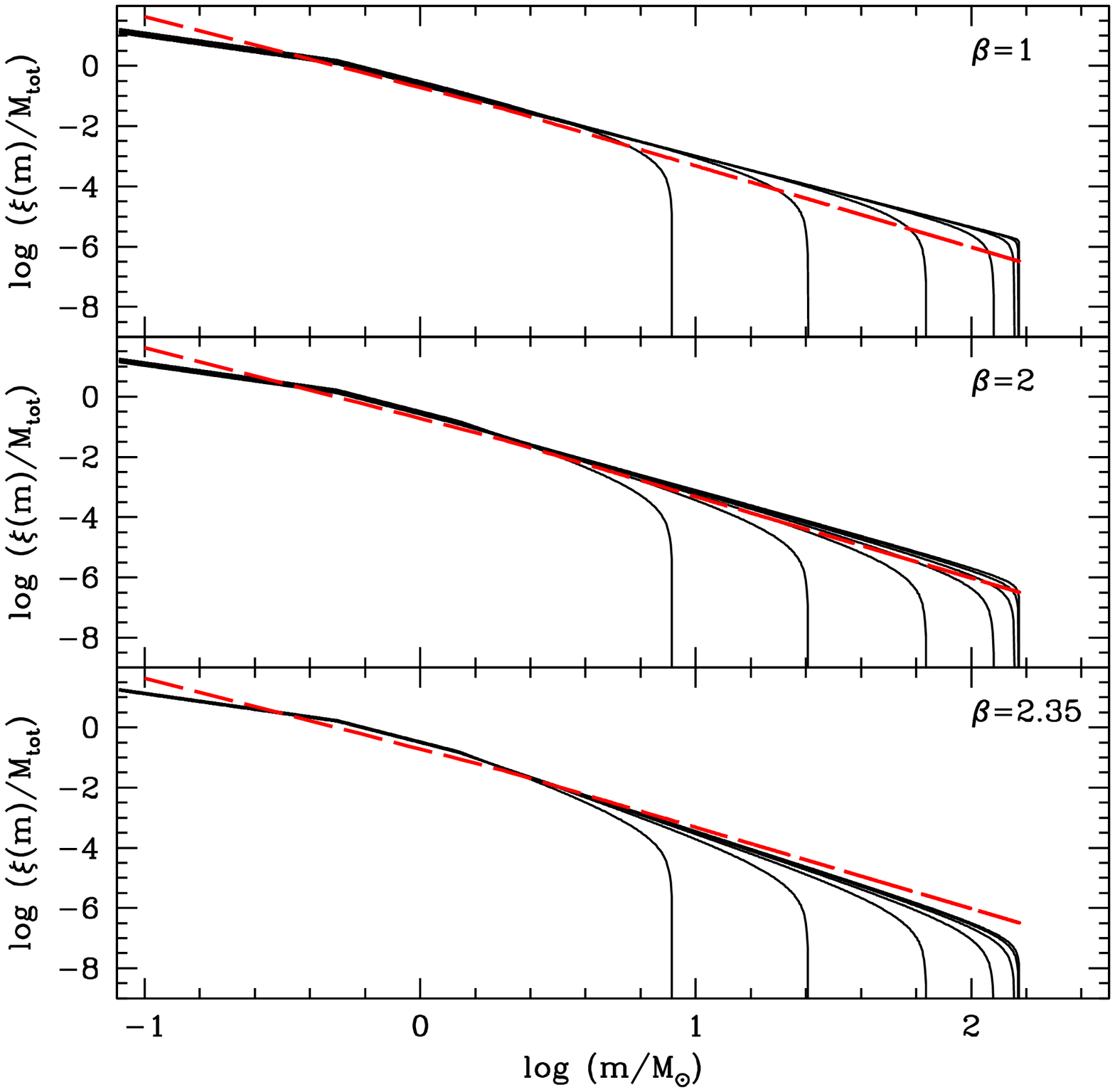}{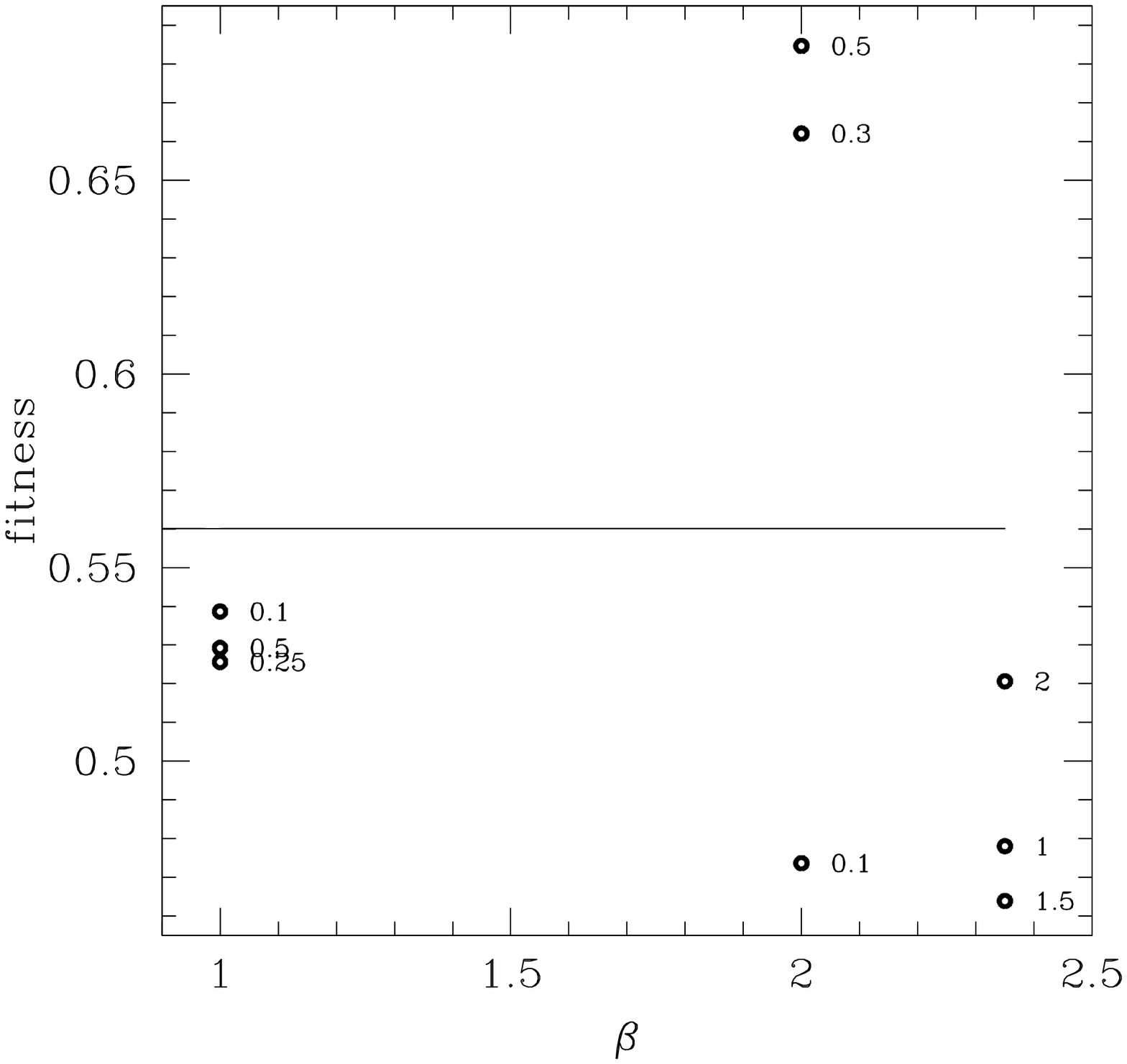}
\caption{\emph{Left panel:} IGIMFs for different cluster mass functions. 
Upper panel: $\beta=1$; central
panel: $\beta=2$; lower panel: $\beta=2.35$. 
In 
each panel we have considered 7 possible values of SFRs, 
ranging
from $10^{-−4}$ $M_{\odot} \, yr^{-1}$ (lowermost solid lines) to 100 $M_{\odot} \, yr^{-1}$ (uppermost
solid lines), equally spaced in logarithm. \emph{Right panel:} the open circles are the ``fitness'' as a function of $\beta$ 
for various models with different SF efficiencies (indicated by the numbers beside each open circle in units of Gyr$^{-1}$.). 
The horizontal line indicates 
the fitness value computed for the standard model. }
\label{fig1}
\end{figure}

\begin{table*}
\vspace{0cm}
\begin{flushleft}
\caption[]{Solar neighbourhood observables, parameters to which they are most sensititve 
and references.}
\begin{tabular}{l|l|l}
\noalign{\smallskip}
\hline
\hline
\noalign{\smallskip}
 Observable              &     Parameter      &          Reference \\ 
\hline
\hline
SFR Surface density & SF efficency                    &  Rana (1991)\\
type Ia SNR         & Integrated SF history, IMF &  Cappellaro  (1996)\\
type II SNR         & SF efficiency, IMF        &  Cappellaro  (1996)\\
Gas surface density & SF efficiency, IMF                 &    Kulkarni \& Heiles (1987)   \\
                    &                                                           &  Olling \& Merrifield (2001)  \\
Stellar surface density          & SF history, IMF            &   Weber \& de Boer (2009)  \\
Stellar abundance ratios &  SF history, IMF & various authors    \\
Stellar Metallicity distribution  & SF history, IMF & Jorgensen (2000)    \\ 
Present-day mass function  & SF history, IMF & Miller \& Scalo (1979)    \\ 
\hline
\hline
\end{tabular}
\label{tab1}
\end{flushleft}
\end{table*}

\section{The (integrated galactic) initial mass function}
\label{sec_igimf} 
The main equation used to calculate the IGIMF is (see the contributions by Pflamm-Altenburg et al. and Weidner et al.):
\begin{equation}
\xi_{\rm IGIMF}(m;{\psi} (t)) = 
\int_{M_{\rm ecl, min}}^{M_{\rm ecl, max} ({\psi} (t))} 
\hspace{-0.6cm}\xi (m \leq m_{\rm max}) \xi_{\rm ecl} (M_{\rm ecl}) 
dM_{\rm ecl}, 
\label{eq1}
\end{equation}
where $\psi$ is the star formation rate (SFR).
The canonical  stellar IMF is $\xi(m) =
k m^{-\alpha}$, with $\alpha = 1.3$ for 0.1 M$_\odot \le$ $m$ $<$ 0.5
M$_\odot$ and $\alpha = 2.35$ for 0.5 M$_\odot \le$
$m$ $< m_{\rm max}$.  The upper mass $m_{\rm max}$ is a function of  the mass of the
embedded cluster $M_{\rm ecl}$: this is logical if one considers that small clusters do not have
enough mass to produce very massive stars. 
Star clusters are also apparently distributed according to
a single-slope power law, $\xi_{\rm ecl} \propto M_{\rm ecl}^{-\beta}$ (Lada \& Lada 2003).  In this work we have assumed 3 possible
values of $\beta$: 1, 2 and 2.35.\\
$M_{\rm ecl, min}$ and $M_{\rm ecl, max} (\psi (t))$ are the
minimum and maximum possible masses of the clusters in a population
of clusters, respectively, and $m_{\rm max} = m_{\rm max} (M_{\rm ecl})$.  
For $M_{\rm ecl,min}$ we take 5 M$_\odot$ (the mass of a Taurus-Auriga aggregate, which is
arguably the smallest star-forming "cluster" known).  The upper mass of the
cluster population depends instead on the SFR and that makes the whole IGIMF dependent on $\psi$.  \\
The standard IMF is a two-slope power law, defined in number as:
\begin{equation}
     \xi_{std}(m) = \left\{ \begin{array}{l l}
                                      0.19\, \cdot m^{-2.35} & 
			     \qquad {\mathrm{if}} \; m < 2 \, M_\odot \\
			              0.24\, \cdot m^{-2.70} &
			     \qquad {\mathrm{if}} \; m > 2 \, M_\odot, \\
                                      \end{array} \right.
\end{equation}

This equation represents  a simplified two-slope approximation of  the actual  
Scalo (1986) IMF. 
The IMF and all the IGIMFs are normalised in mass  to unity 
as the standard IMF:
\begin{equation}
\int m \xi (m) dm = 1.
\label{norm}
\end{equation}

In Fig.~\ref{fig1}, we show the IGIMF as a function of the SFR for the three values of $\beta$ considered 
in this work, compared to our standard IMF.
In general, a lower value of $\beta$ implies a flatter IGIMF, 
and a hence higher relative fraction  of 
stars with masses $m>1M_{\odot}$.

\begin{figure}
\plotthree{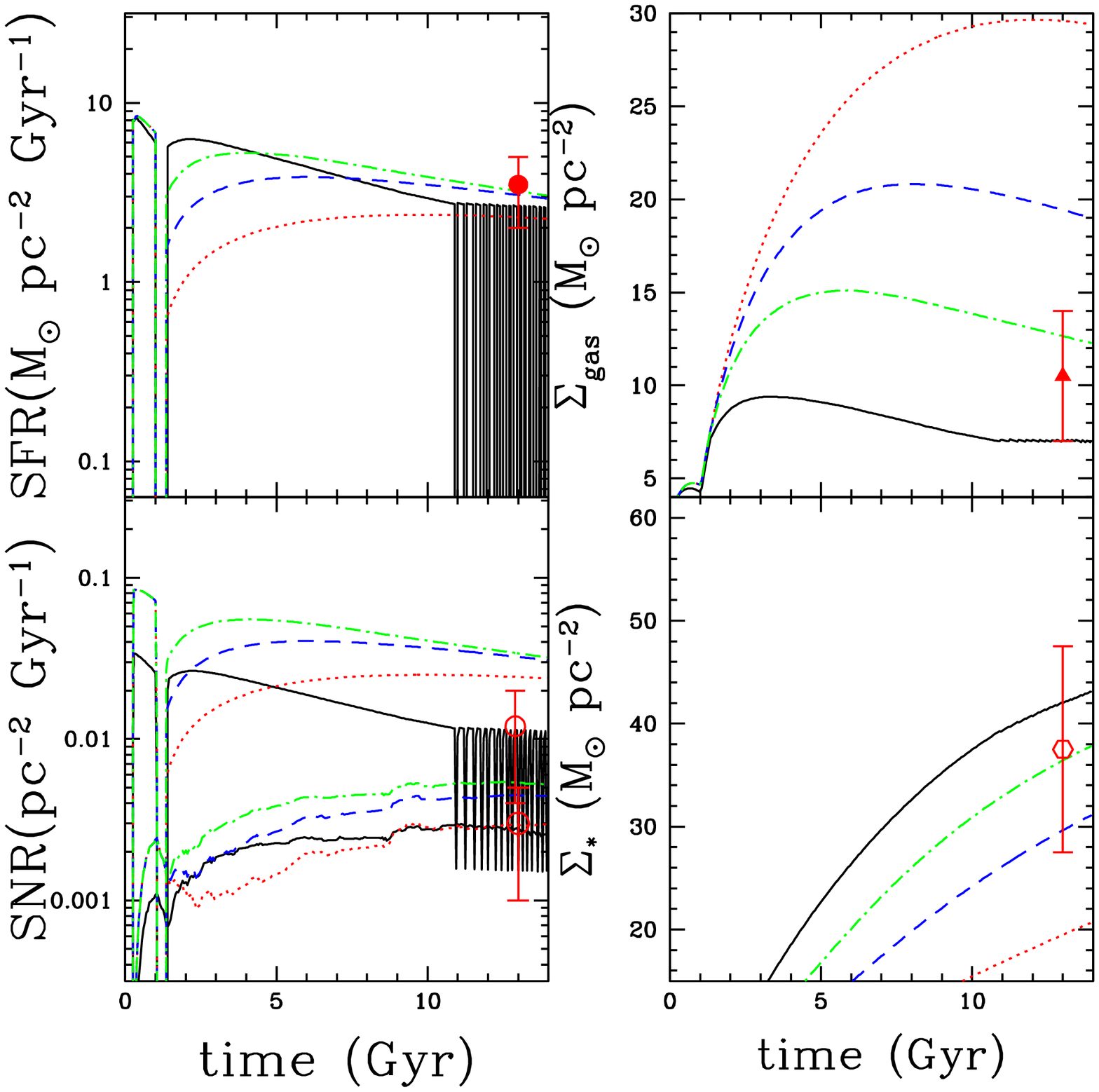}{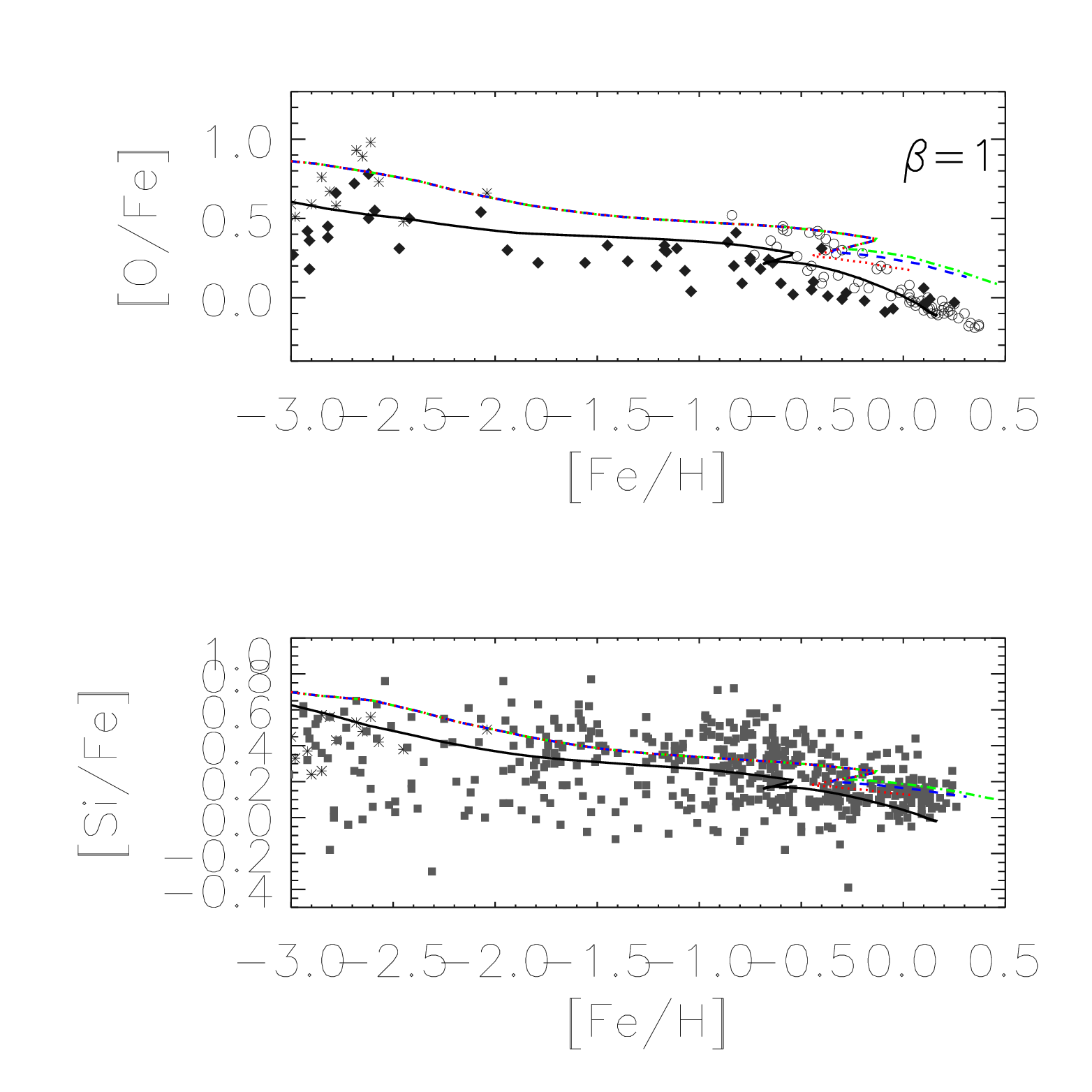}{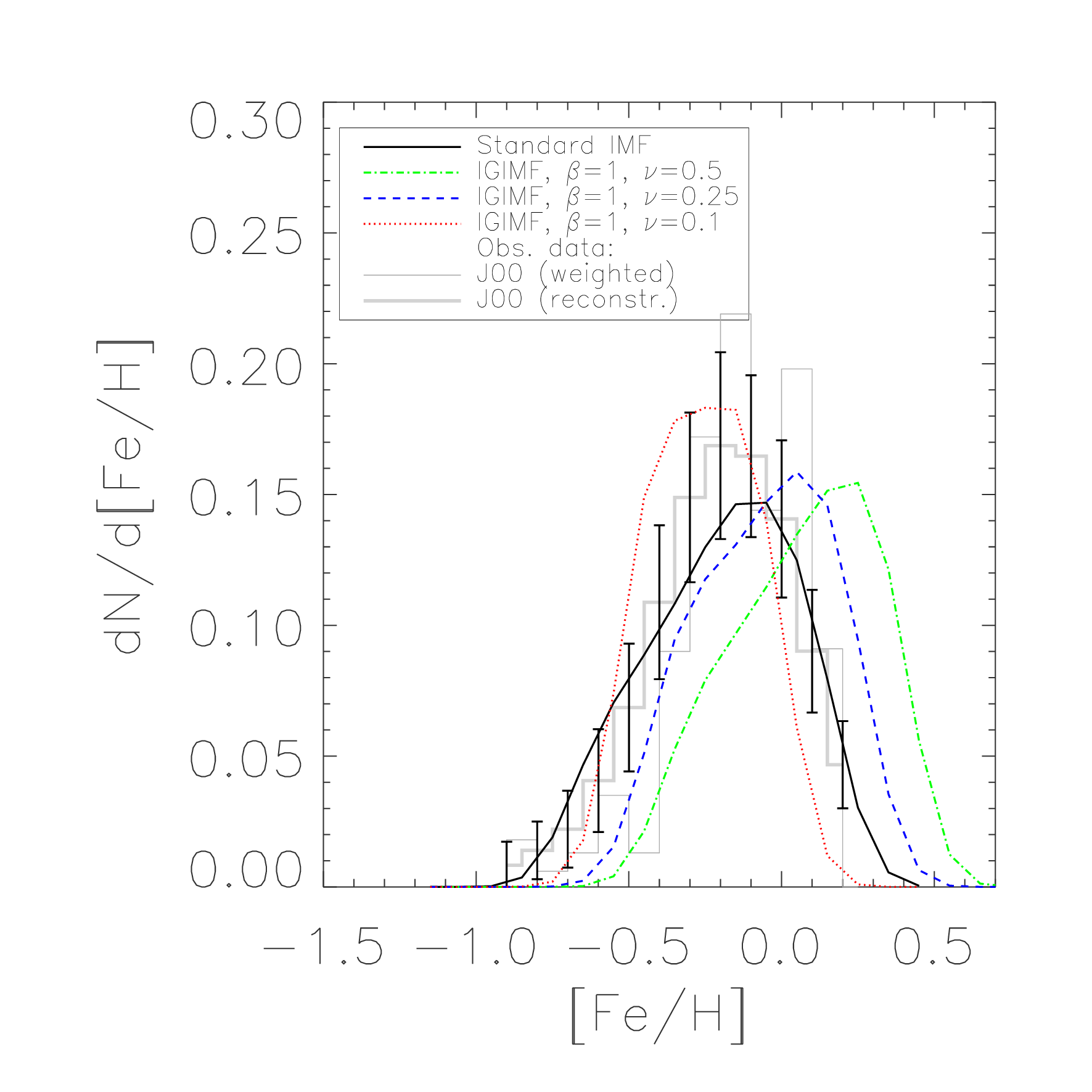}
\plotthree{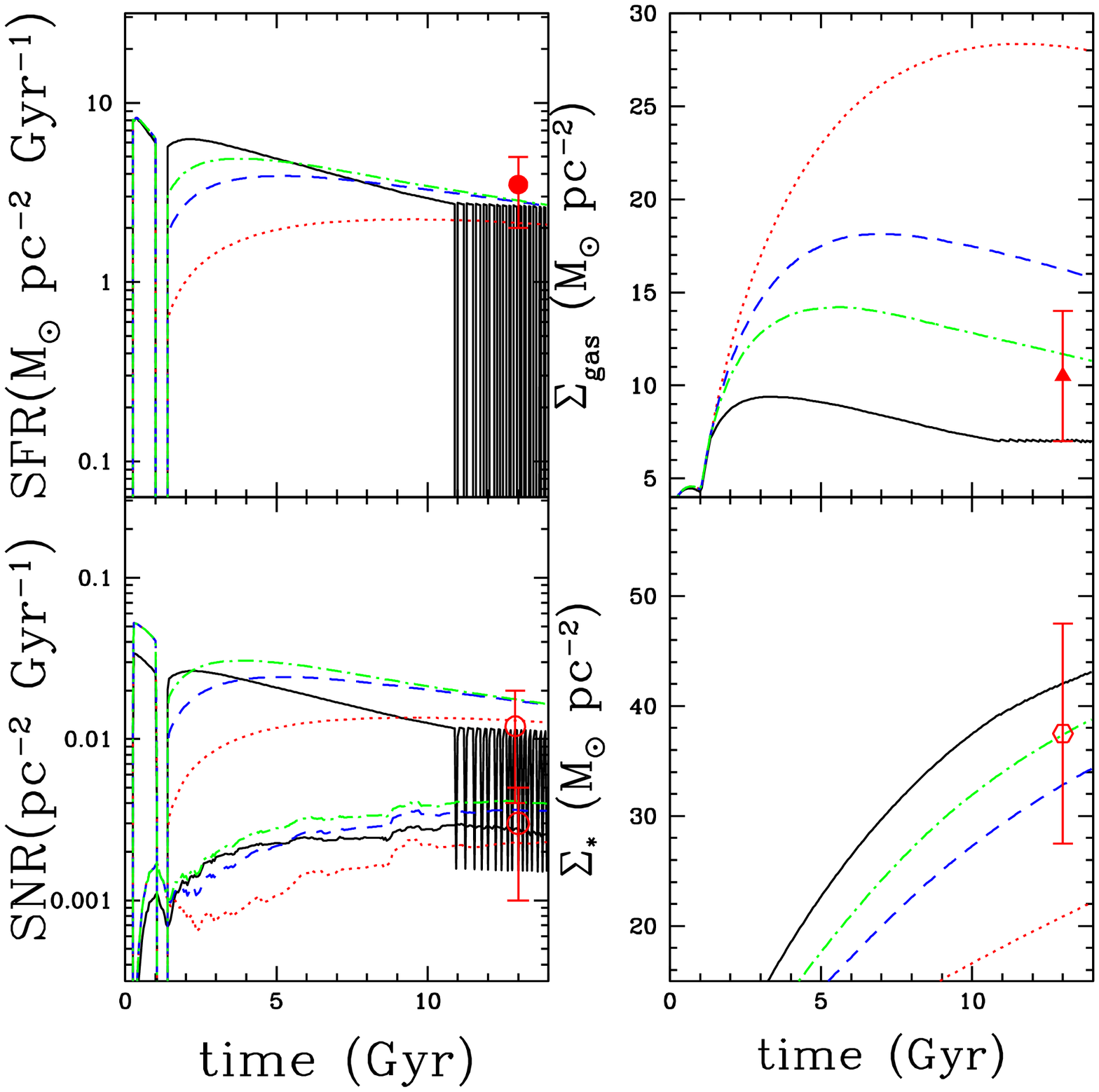}{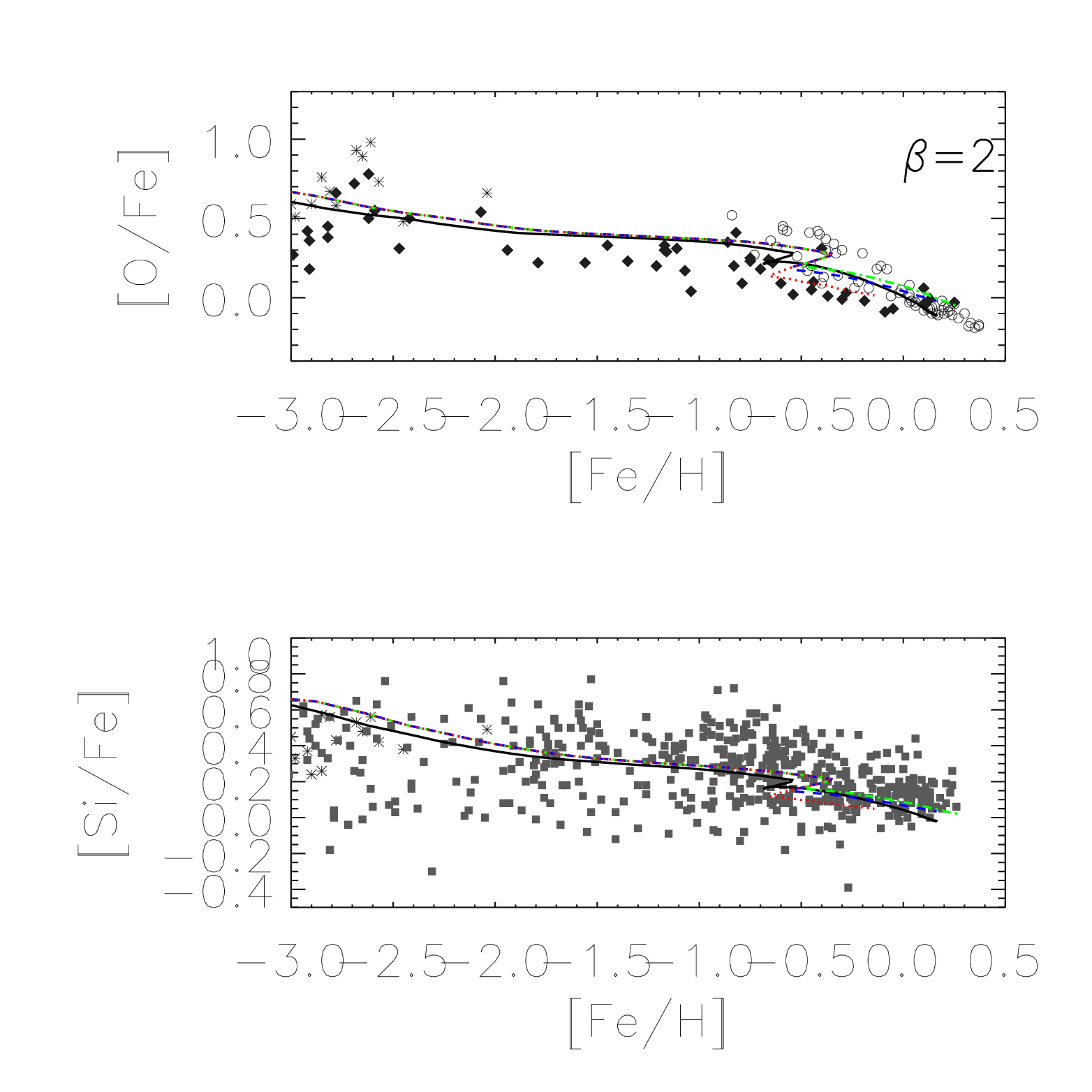}{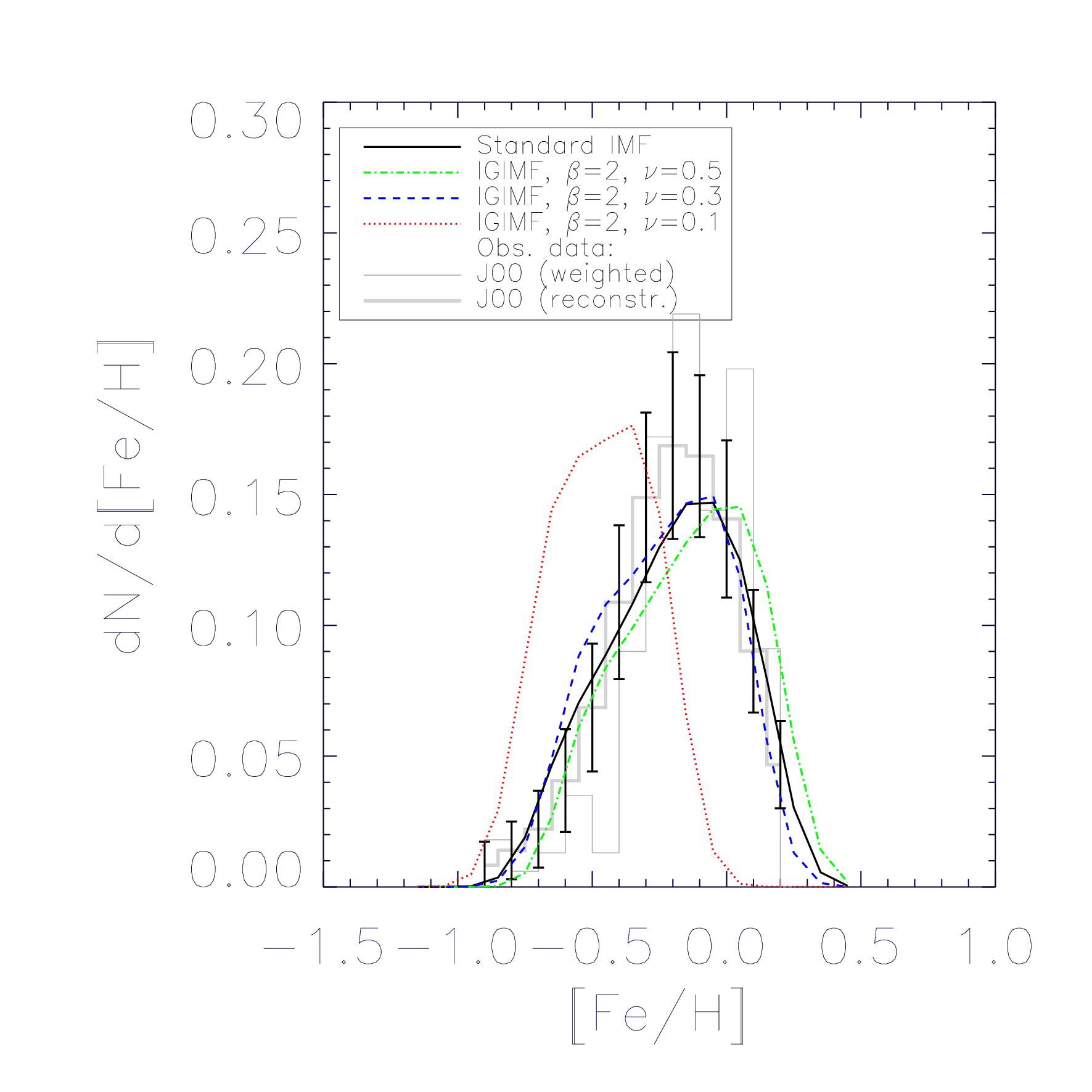}
\plotthree{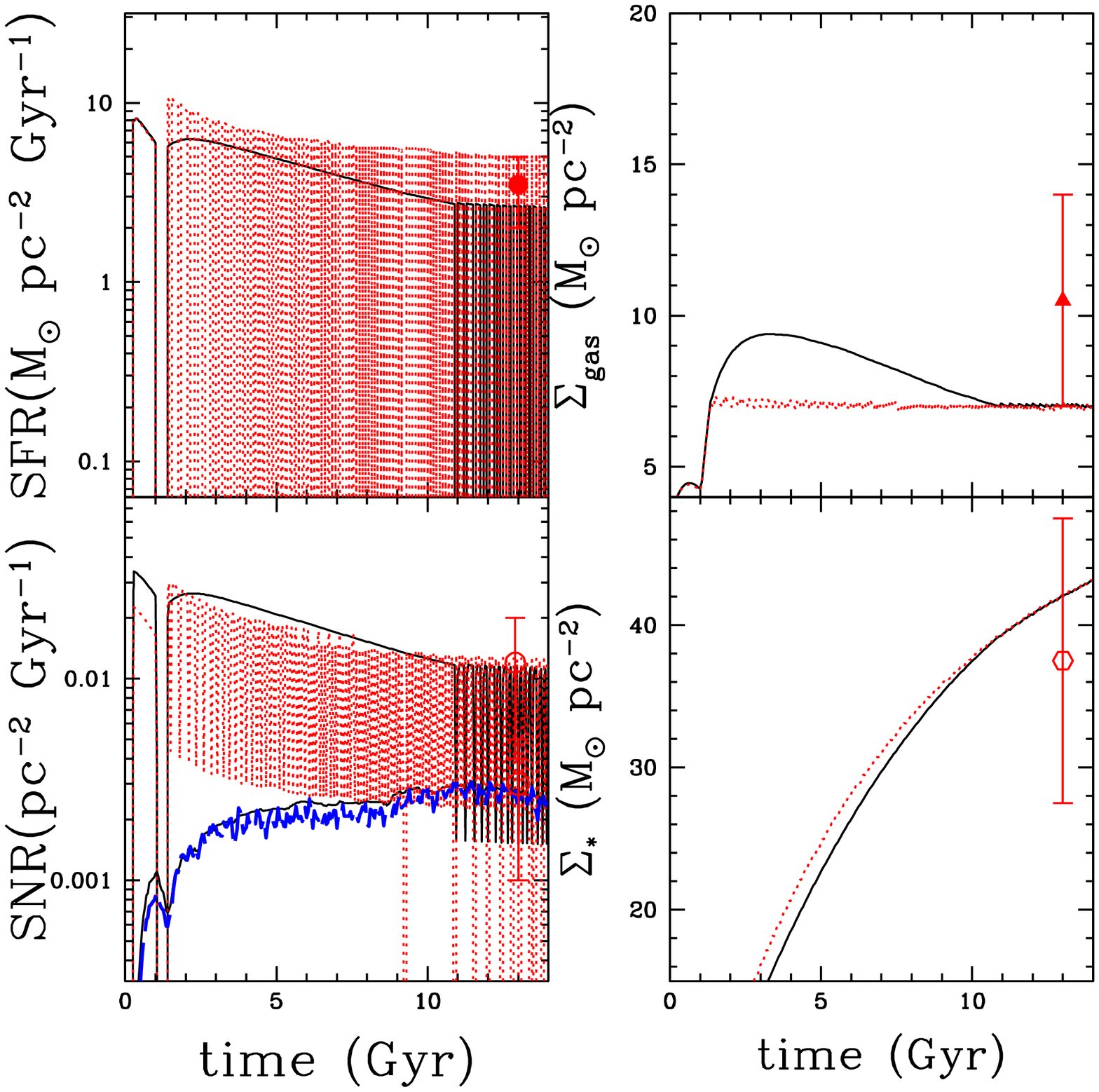}{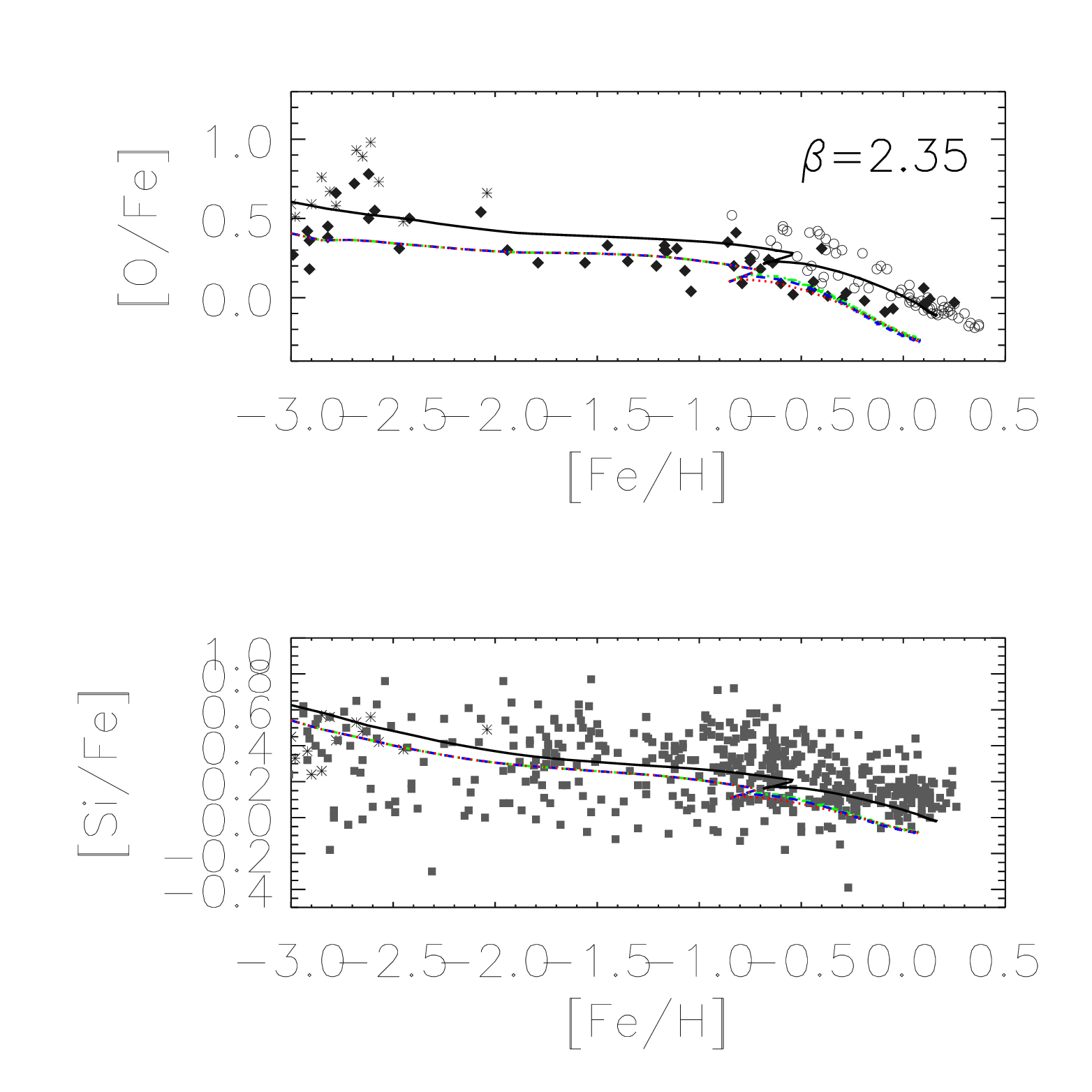}{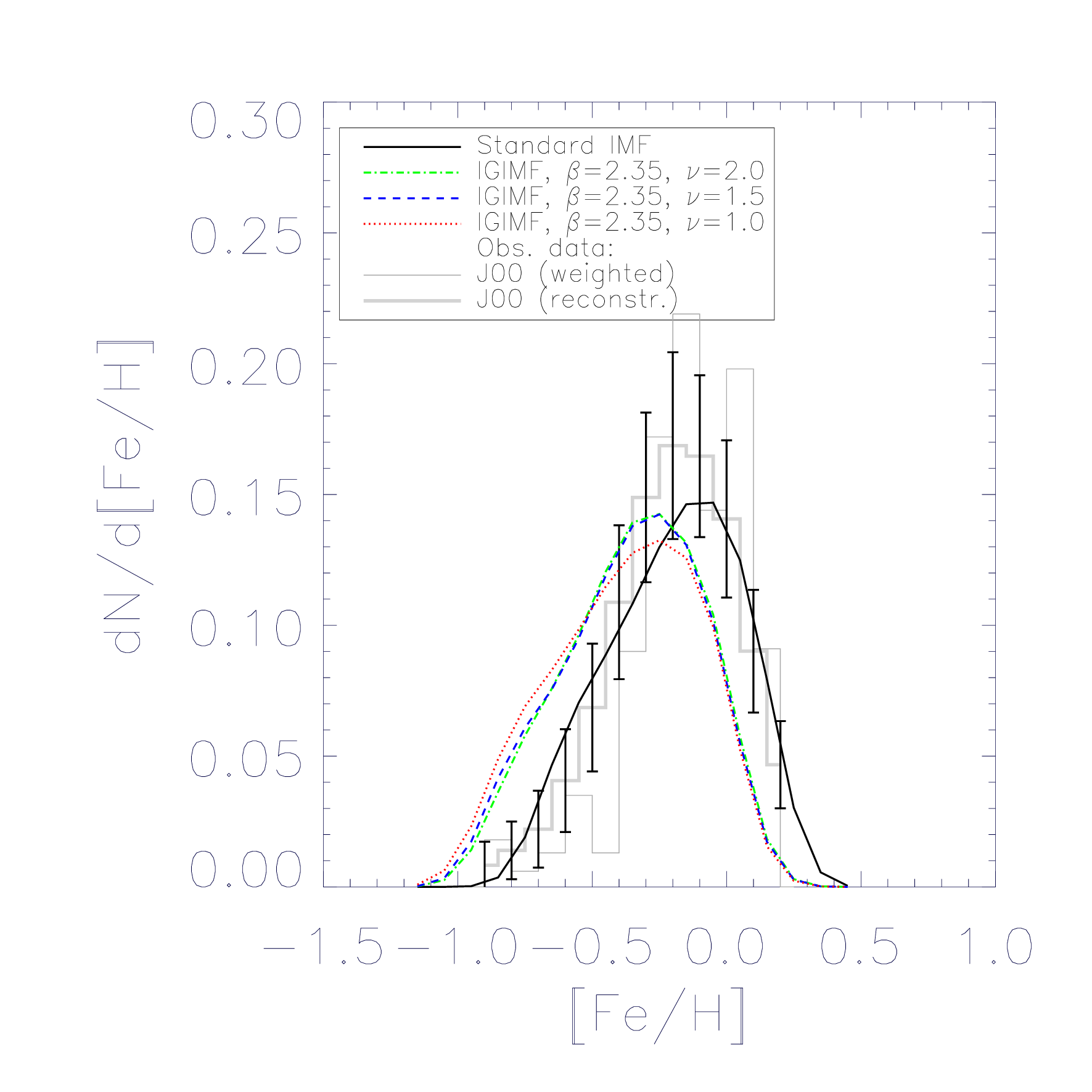}
\caption{\emph{Upper panels}: SFR, type Ia and II SN rates, stellar and gas density versus time (left),
 abundance ratios (center) and stellar metallicity distribution (right) calculated with $\beta=1$. \emph{Middle panels}: same as above but 
with $\beta=2$. \emph{Lower panels}: same as above but 
with $\beta=2.35$ (in the left panels only the case with SF efficiency $1.5$ Gyr$^{-1}$ is shown). For details see Calura et al. (2010).}
\label{fig2}
\end{figure}

\section{The chemical evolution model for the Solar Neighbourhood}
\label{sec_gce}
The main chemical evolution model is described in detail in Calura et al. (2010). 
The  model is calibrated in order to reproduce a large set of observational 
constraints for the Milky Way galaxy (Chiappini et al. 2001). 
The Galactic disc is approximated by several independent rings, 
2 kpc wide, without exchange of matter between them. The Milky Way 
is  assumed to form as a result of two main infall episodes.  
During the first episode, the halo and the thick disc are formed.
During the second episode, a slower infall
of external gas forms the thin disc with the gas accumulating faster in the inner than in the outer
region ("inside-out" scenario, Matteucci \& Fran\c cois 1989). The process of disc formation is much longer than the
halo 
and bulge formation, with time scales varying from $\sim2$ Gyr in the inner disc to $\sim7$ Gyr in the solar region
and up to $20$ Gyr in the outermost disc. 
In this paper, we are interested in the effects of a time-variyng IMF in the Solar Neighbourhood. 
For this purpose, we focus on a ring located at 8 kpc from the Galactic centre, 2 kpc wide. 
The model includes the contributions of type Ia SNe, type II SNe and low and intermediate mass stars to the chemical 
enrichment of the ISM. 
A SF threshold is adopted ($\sigma_{th}\sim 7 M_{\odot} pc^{-2}$) to reproduce various local features, including abundance gradients (Colavitti et al. 2009). 
The IGIMF is allowed to vary as a function of the SFR, 
which in turn is a function of  cosmic time. The IGIMF is calculated as a function of the SFR 
according to Eq.~\ref{eq1}. 
In Table ~\ref{tab1} we show the solar neighbourhood observables used in this paper, with the main parameters on which they depend and references.\\
In order to quantitatively compare our results with the observables
considered in this paper, we define the $fitness$ quantity as:
\begin{equation}
fitness = {1 \over {1 + \delta}}~~;~~\delta = \Sigma_i {w(i)
[{\rm obs} (i) - {\rm theo} (i)]^2 \over 
\max\bigl\lbrace[{\rm obs} (i)]^2, [{\rm theo} (i)]^2\bigr\rbrace}
\end{equation}
\noindent
where, for the $i$-th value of each considered parameter, obs ($i$) and theo ($i$)
are the observed values and the predictions of the model, respectively.  
The weight $w (i)$ is used to give
each set of observables the same statistical weight.\\
The closer $fitness$ is to 1, the better the
model is in reproducing the observations.
In the right panel of Fig.~\ref{fig1}, we show 
the ``fitness'' as 
a function of $\beta$ for all the models considered in this paper (see Sect. \ref{res}).

\section{Results}
\label{res}
Some results of our study are summarized in Fig.~\ref{fig2}. 
Here we show the results of the comparison between observables and model predictions computed with different assumptions for the 
cluster mass function index $\beta$. In general, the local 
observed quantities are plotted with their error bars. For a given value of $\beta$, 
the curves of different types are the model results computed with different assumptions 
for the SF effiency, reported in the legends. 
From the time evolution of the SFR, type Ia and II supernova rate (SNR), gas and stellar mass density (left panels) computed with 
various assumptions for the parameter $\beta$, we can see that lower values of $\beta$ 
imply higher SNRs, higher gas mass densities and in general lower 
mass locked up in living stars and remnants. This is visible from the comparison of the results computed with the same 
values for the SF efficiencies and different values for $\beta$. Another important issue regards the dependence on the star formation 
threshold: some values of  $\beta$ ($\beta=1$,  $\beta=2$) show SF histories indendent from the SF threshold. 
This is basically due to the large mass returned by dying stars, which maintains the gas density always above the threshold level 
and which produces SF histories substantially  
different from those obtained with the standard IMF, for which the effect of the threshold is remarkable, in particualr in the last 4 Gyr  
of evolution. On the other hand, the case with $\beta=2.35$ shows a dependence on the SF threshold even larger than the standard case, 
and this is due to the lower mass ejection rates from dying stars which stem from a steeper IGIMF (see Fig. ~\ref{fig1}). \\
In the middle panel, we show the abundance pattern predicted for the three values of $\beta$, compared to those predicted in the standard 
case. The elements studied here are O,Si and Fe, since the theoretical understanding of their production is quite 
clear, and their study  allows us 
to neglect uncertainties related to their nucleosynthesis. 
In general, lower values of $\beta$ produce higher [Fe/H] values  and higher  
[$\alpha$/Fe] values (i.e. more $\alpha$-enhanced elemental abundances)  at a given metallicity. This is related to 
the fact that the fraction of massive stars increases with decreasing $\beta$  and considering that  massive stars are the main producers of  $\alpha$ elements. 
Similar conclusions can be drawn by looking at the plots of the stellar metallicity distributions: the larger the 
value of $\beta$, the larger the relative fraction of stars producing Fe, i.e. mostly type Ia SNe, i.e. stars 
in binary systems with initial mass ranging from $0.8 M_{\odot}$ to $8 M_{\odot}$, hence larger the 
Fe abundances at any given epoch. This translates in SMDs peaking at higher [Fe/H] values for lower values of 
$\beta$, assuming the same SF efficiency. \\
As can be seen from the right panel of Fig.~\ref{fig1}, the model calculated with the IGIMF providing the best results is the one 
with $\beta=2$ and SF efficiency 0.3-0.5 Gyr$^{-1}$. 
The results  obtained with this choice of $\beta$ are quite similar to the ones achieved with the standard IMF. 
This should not be a surprise  since, as shown in Fig.~\ref{fig1}, 
in the intermediate case with $\beta=2$  the IGIMF is very similar to the standard IMF.  
The assumption of $\beta=2$ allows us to satisfactorily reproduce the set of observational constraints considered in this work. 
This is an important result, given the fact 
that the 
IGIMF is computed from first principles. 
On the basis of the results described in this 
section, it may be difficut to discriminate between the scenario with the standard IMF and 
the IGIMF with $\beta=2$. 
In the next section, the use of a diagnostic possibly useful to disentangle between the standard IMF 
and the IGIMF will be discussed.

\section{Discussion}

In this paper, 
We have modelled the physical properties of the S. N. within the IGIMF theory. 
In this scenario, the IGIMF can be calculated by combining the cluster mass function with the stellar IMF, which represents the 
mass function of stars born within clusters and which 
can be described by a double-slope power law. 
An  important feature of the IGIMF is that it depends on the star formation rate, which in turn evolves with time. 

The parameter $\beta$ regulating  the cluster mass function may have an important impact on the 
predicted properties of the Solar Neighbourhood. 
In general alower value for $\beta$ corresponds to  a flatter IGIMF. 
In terms of chemical evolution, a flatter IGIMF translates into 
higher mass ejection
rates from dying stars, hence globally a lower mass fraction incorporated into stellar remnants and higher gas mass densities.
This implies that the evolution of all the models computed assuming $\beta=1$ and $\beta=2$ are not sensitive to the star formation 
threshold and the star formation histories do not exhibit the ``gasping'' features typical of the standard model, which in turn is dominated by 
threshold effects at evolutionary times greater than 10 Gyr. 
Moreover, the lower the value of $\beta$, the higher the SN rate, the higher the metallicity and the larger the $\alpha$-enhancement 
visible in the abundance pattern. 
The statistical test used to compare model results and the obervables 
indicates that the model which best reproduces the local observables is carachterized by $\beta=2$ as the index of the CMF. 
The results of the best model are very similar to those obtained with the standard case. A possible diagnostic which could help us disentangling between the two 
is represented by the present-day mass function (PDMF). 
The PDMF represents 
the mass function of living stars as observed in the solar neighbourhood. This quantity  
is an important diagnostic since 
it provides pieces of information complementary to the ones from the previously discussed observables. 

\begin{figure}
\plottwo{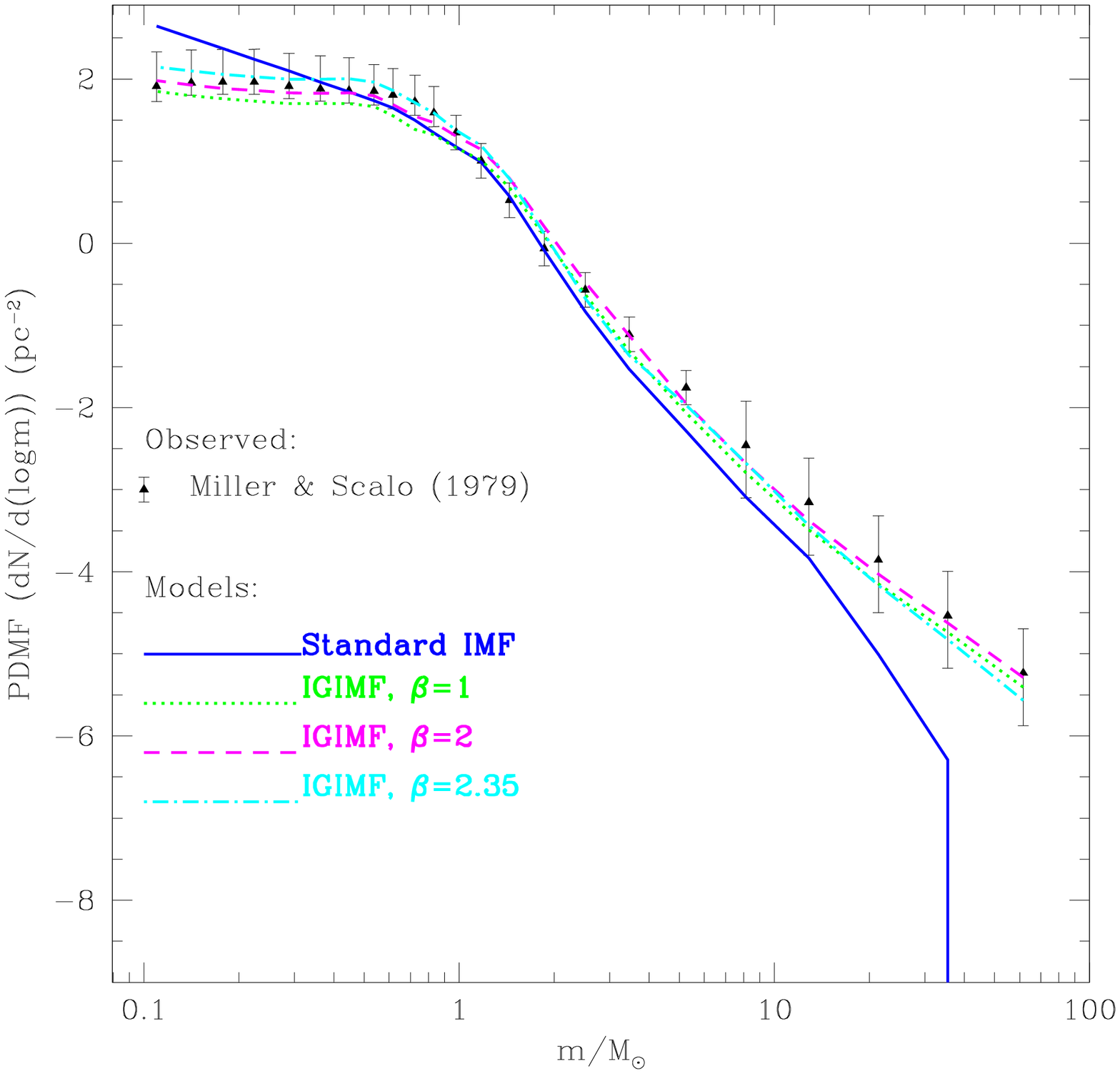}{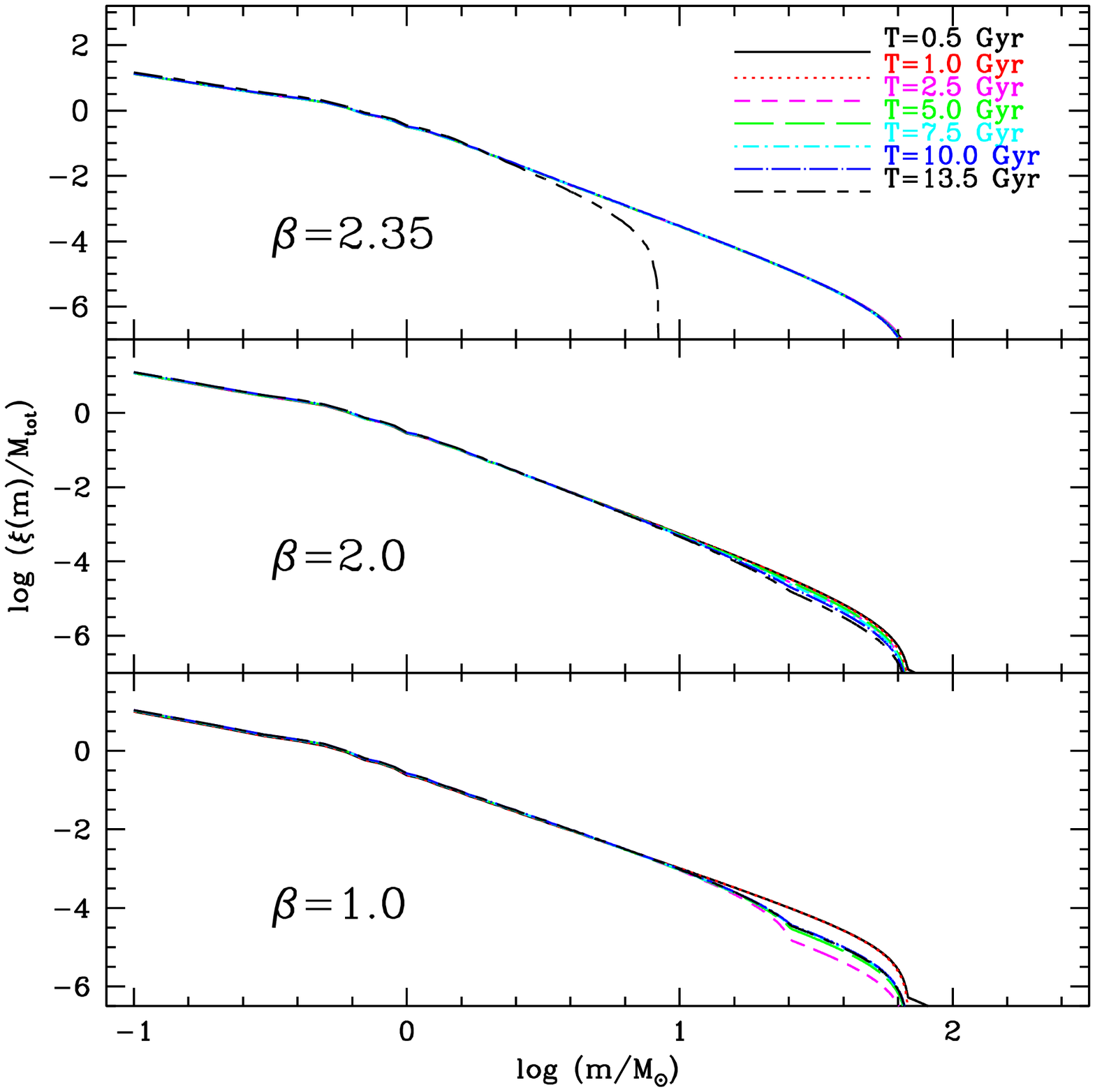}
\caption{\emph{Left panel:} present-day mass function as predicted by means of our models compared to local observations.  \emph{Right panel:} time evolution of the IGIMF 
assuming various values for the parameter $\beta$.}
\label{fig3}
\end{figure}

In the left panel of Fig.~\ref{fig3}, we show the  PDMF observed in the S. N.  
and predicted by means of our models. 
The PDMF computed with the standard IMF agrees with the observations in the range 0.4 $ M_{\odot}$  - 2  $M_{\odot}$. 
At very low stellar masses, the standard IMF seems  too steep, whereas the 
distribution of stars with masses $>30  M_{\odot}$ is underestimated. 
Once again, this is due to the SF threshold, which has 
strong effects on the SF history of the solar neighbourhood at late times, inhibiting recent SF and hence causing the underabundance or absence 
of very massive stars. In contrast,  
the models calculated with the IGIMF provide all similarly  
a very good fit to the observed PDMF. 
\\
The analysis of Fig.~\ref{fig3} seems to suggest that the SF threshold should not play a dominant 
role in the late evolution of the S. N. 
Within the IGIMF theory, the  existence of a SF threshold may 
be an observational selection effect, naturally explained in this context as shown by Pflamm-Altenburg et al. (these proceedings). 
However, as shown by chemical evolution results, the SF threshold is fundamental in reproducing 
the metallicity gradients observed in the MW and in local galaxies, unless a variable star formation efficiency through the disc is assumed 
(Colavitti et al. 2009).  The study of the abundance gradients within the IGIMF theory may be of crucial help in sheding light on this issue 
and will be considered in future work. \\
Another important issue concerns the time evolution of the IGIMF. 
In the right panel of Fig.~\ref{fig3}, we show how the IGIMF varies as  a function of time 
in the case of the three best models computed with different values of $\beta$. 
The best model ($\beta=2$) shows 
very small variation of the IGIMF with cosmic time. Strong variations are predicted by assuming  $\beta=2.35$, 
since in this case the star formation history is very much influenced by the effects of the SF threshold.
It may be interesting to test the effects of the IGIMF in spiral, Milky Way-like galaxies in a cosmological context. 
Cosmological semianalytical  models predic strong variations in the star formation histories of spiral galaxies (Calura \& Menci 2009), which 
present a large number of spikes due to merging events and which 
should manifest into strong variations of the IGIMF with redshift. \\
The universality of the IGIMF is another issue that deserves particular attention in the future. A chemical evolution study of 
elliptical galaxies within the IGIMF theory shows that the best value for $\beta$ is 2.35, allowing to reproduce  best the integrated 
$\alpha/Fe$ ratios observed in the local early-type galaxies. This value is in contrast with the best value suggested by the analysis of 
the S. N. features. A further study of the IGIMF in local dwarf irregular galaxies and dwarf spheroidals could certainly be of some help in this regard. \\
Currently, 
the slope of the IGIMF in extreme SF conditions is another largely debated topic. 
Various indirect indications in external galaxies (Dabringhausen, these proceedings) and in our Galaxy (Stolte, these proceedings) 
seem to suggest that in strongly star forming systems, the slope of the stellar IMF should be flatter than the Salpeter one. 
Moreover, the assumption of a slightly top-heavy IMF in starbursts helps alleviating the discrepancy between cosmological models and observations 
regarding the $\alpha/Fe$-$\sigma$ relation observed in local ellipticals (Calura \& Menci 2009). 
Addressing this subject within the IGIMF theory will be of primary importance in the nearest future.

\acknowledgements FC would like to thank the S.O.C. for the kind invitation and for the financial support, whereas the L.O.C. is acknowledged for 
being able to establish a pleasant and comfortable environment to discuss exciting scientific topics.

\bibliography{calura_f}


\end{document}